\documentclass[prd,twocolumn,floatfix,superscriptaddress]{revtex4}
\pdfoutput=1
\usepackage{graphicx}
\usepackage{epsfig}
\usepackage{bm}
\usepackage{amssymb}
\usepackage{float}
\usepackage{amsmath}
\usepackage{dcolumn}
\usepackage{cancel}
\usepackage[colorlinks]{hyperref}
\usepackage[usenames,dvipsnames]{color}
\usepackage{subfigure}
\hypersetup{
     breaklinks=true,
                        pdfstartview={FitH},                                    
colorlinks=true,    
       linkcolor=blue,              citecolor=red,            filecolor=magenta,          
urlcolor=b
lue,               anchorcolor=green,          linktocpage=true
}

\def\doi{http://doi.org}

\begin{document}

\title{Dynamical system analysis at background and perturbation levels: 
Quintessence in severe 
disadvantage comparing to $\Lambda$CDM  }
\author{Spyros Basilakos}
\affiliation{Academy of Athens, Research Center for Astronomy and Applied Mathematics,
Soranou Efesiou 4, 11527, Athens, Greece}
\affiliation{National Observatory of Athens, V. Paulou and I. Metaxa 15236, Penteli, 
Greece}
\author{Genly Leon}
\affiliation{Departamento de Matem\'aticas, Universidad Cat\'olica del Norte, Avda.
Angamos 0610, Casilla 1280 Antofagasta, Chile}
\author{G. Papagiannopoulos}
\affiliation{Physics Department,
University of Athens, Panepistemiopolis, Athens 157 83, Greece}
\author{Emmanuel N. Saridakis}
\affiliation{Physics Division, National Technical University of Athens, 15780 Zografou
Campus, Athens, Greece}
\affiliation{Department of Astronomy, School of Physical Sciences, University of Science
and Technology of China, Hefei 230026, China}

\begin{abstract}
We perform for the first time a dynamical system 
analysis of both the background and 
perturbation equations, of $\Lambda$CDM  cosmology and quintessence 
scenario with an 
exponential potential. 
In the former case the
perturbations do not change the stability of the late-time 
attractor of the background equations, and the system still results in the dark-energy 
dominated, de Sitter solution, having passed from the correct dark-matter era with 
$\gamma\approx6/11$. However, in the case of quintessence the incorporation of 
perturbations changes the 
stability and properties of the background evolution, and the only conditionally stable 
points present either  an exponentially increasing matter 
clustering not favored by observations, or Laplacian instabilities, and thus   not 
physically interesting. This result is a severe disadvantage of 
quintessence cosmology comparing to $\Lambda$CDM paradigm.

\end{abstract}

%\pacs{98.80.-k, 95.36.+x, 04.50.Kd}
\maketitle

%\email{svasil@academyofathens.gr}

%\email{genly.leon@ucn.cl}

%\email{yiannis.papayiannopoulos@gmail.com}

%\email{Emmanuel\_Saridakis@baylor.edu}

\section{Introduction} 
Dynamical system approach is a powerful tool that
allows to extract information on the evolution of a cosmological model,
independently of the initial conditions or its specific behavior at
intermediate times \cite{REZA}. In particular, although a general
cosmological scenario may exhibit an infinite number of possible evolutions,
its asymptotic behavior, namely its behavior at late times, can be
classified in a few different classes, which correspond to the stable
critical points of the autonomous-form transformed cosmological equations.
Thus, through such an analysis one obtains information of the late-time
universe, bypassing the complications of the cosmological equations, which
prevent complete analytical treatments, as well as the ambiguity of the
initial conditions.

The dynamical system approach has been applied to numerous cosmological
scenarios since the late 90's (see \cite{Bahamonde:2017ize} and references
therein), nevertheless up to now it remained only 
at the background level, namely
examining the behavior of the background equations and calculating at the
critical points the values of background-related quantities such as the
density parameters, the equation-of-state parameter etc. Although this
analysis was important and adequate for the earlier cosmology advance, the
significantly advancing cosmological progresses and especially the huge
amount of data related to perturbations (such as the growth index and the
Large Scale Structure), leads to the need to extend the dynamical system
approach in order to investigate cosmological scenarios at both the
background and perturbation levels.  \newline
\begin{table*}[ht]
\begin{center}
\begin{tabular}{|c|c|c|c|c|c|c|}
\hline
C.P. & $x$ & $y$ & Existence & Stability & $\Omega_d$ & $w_d$ \\ \hline\hline
A & 0 & 0 & Always & Saddle for $0 < \gamma_m < 2$ & 0 & Undefined \\ \hline
B & 1 & 0 & Always & Unstable node for $\lambda < \sqrt{6}$ & 1 & 1 \\ 
&  &  &  & Saddle for $\lambda > \sqrt{6}$ &  &  \\ \hline
C & -1 & 0 & Always & Unstable node for $\lambda > -\sqrt{6}$ & 1 & 1 \\ 
&  &  &  & Saddle for $\lambda < -\sqrt{6}$ &  &  \\ \hline
D & $\lambda/\sqrt{6}$ & $[1-\lambda^2/6]^{1/2}$ & $\lambda^2 < 6$ & Stable
node for $\lambda^2 < 3\gamma_m$ & 1 & $\frac{\lambda^2}{3}-1$ \\ 
&  &  &  & Saddle for $3\gamma_m < \lambda^2 < 6$ &  &  \\ \hline
E & $(3/2)^{1/2} \, \gamma_m/\lambda$ & $[3(2-\gamma_m)\gamma_m/2%
\lambda^2]^{1/2}$ & $\lambda^2 > 3\gamma_m$ & Stable node for $3\gamma_m <
\lambda^2 < 24 \gamma_m^2/(9\gamma_m -2)$ & $3\gamma_m/\lambda^2$ & $w_m$ \\ 
&  &  &  & Stable spiral for $\lambda^2 > 24 \gamma_m^2/(9\gamma_m -2)$ &  & 
\\ \hline
\end{tabular}%
\end{center}
\caption[Quintback]{The critical points, their stability conditions (the
corresponding eigenvalues are given in \protect\cite{Copeland:1997et}), and
the values of $\Omega_{d}$ and $w_{d}$, for the quintessence scenario with 
exponential potential, with $%
\protect\gamma_m\equiv w_m+1$. }
\label{Quintback}
\end{table*}

\section{Dynamical analysis at the background level} 
Let us briefly review
the phase space analysis of $\Lambda$CDM paradigm, as well as of the basic
dynamical dark energy scenario, namely the quintessence one with an exponential 
potential, which is the archetype quintessence scenario due to the well-posed theoretical 
justification of exponential potentials. Considering a
flat Friedmann-Robertson-Walker (FRW) metric $ds^2= dt^2-a^2(t)\,
\delta_{ij} dx^i dx^j$, the equations of a general cosmological scenario
read as 
\begin{eqnarray}
&&H^{2}=\frac{\kappa^2}{3}(\rho_m+\rho_{d}),  \label{FR1} \\
&&\dot{H}=-\frac{\kappa^2}{2} (\rho_m+p_m+\rho_{d}+p_{d}),  \label{FR2}
\end{eqnarray}
with $\kappa^2=8\pi G$, and where $\rho_m$, $p_m$ are respectively the energy
density and pressure of the matter fluid, while $\rho_{d}$, $p_{d}$ are the
energy density and pressure of the (effective) dark energy fluid. Finally,
assuming that interactions do not take place among the cosmic fluid
components, the system of equations closes with the conservation equations 
\begin{eqnarray}
&&\dot{\rho}_m+3H(1+w_m) \rho_m=0,  \label{eqn:contmt} \\
&&\dot{\rho}_{d}+3H(1+w_d) \rho_d=0,  \label{eqn:contdt}
\end{eqnarray}
where we have introduced the equation-of-state parameters $w_i\equiv
p_i/\rho_i$. Note that only three out of four equations (\ref{FR1})-(\ref%
{eqn:contdt}) are independent.

The above framework provides $\Lambda$CDM cosmology for $\rho_d=-p_d=%
\Lambda/\kappa^2$, with $\Lambda$ the cosmological constant, and in this
case Eq. (\ref{eqn:contdt}) becomes trivial. Additionally, for the case of
the basic quintessence scenario, in which a scalar field $\phi$ is
introduced, we have $\rho_d=\dot{\phi}^2/2+V$ and $p_d=\dot{\phi}^2/2-V$,
with $V(\phi)$ its potential, and then Eq. (\ref{eqn:contdt}) becomes the
Klein-Gordon equation $\ddot{\phi}+3 H\dot{\phi}+V^{\prime }=0$, with $%
V^{\prime }(\phi)\equiv \partial V/\partial\phi $.

The essence of the dynamical system approach is to transform the equations
into an autonomous system, using $\tau\equiv\ln a$ as the dynamical
variable, extract its critical points, perturbing around them, and
investigate their stability by examining the eigenvalues of the involved
perturbation matrix \cite{REZA,Bahamonde:2017ize}.

For $\Lambda$CDM cosmology the cosmological equations can be transformed
into an autonomous form by simply using the matter density parameter $%
\Omega_m\equiv \kappa^2\rho_m/(3H^2)$ as the auxiliary variable. Thus, Eqs. (%
\ref{FR1}) and (\ref{eqn:contmt}) give rise to the one-dimensional system 
\begin{equation}
\Omega_m^{\prime }=3(\Omega_m-1)\Omega_m,
\end{equation}
where primes denote derivatives with respect to $\tau$. The system has two
critical points, characterized by $\Omega_m=1$ and $\Omega_m=0$, and one can
see that the former is unstable while the latter stable. Therefore, for $%
\Lambda$CDM cosmology, the cosmological-constant dominated ($\Omega_m=0$
according to (\ref{FR1}) implies that $\Omega_d\equiv (\kappa^2\rho_d/3H^2)=1
$), de-Sitter solution is the stable late-time attractor, and thus the
universe will result to it independently of the initial conditions and its
evolutions at intermediate times. We mention that actually the dynamical
system analysis is not needed in this scenario, since the equations are
integrable, with the solution 
\begin{equation}
\Omega_m=\frac{\Omega_{m0}}{e^{3(1+w_m)\tau}(1-\Omega_{m0})+\Omega_{m0}},
\end{equation}
with $\Omega_{m0}$ the value of $\Omega_{m}$ at $a=1$. Hence, we can
immediately see that at late times the system always reaches the de-Sitter
solution (for matter sectors that do not violate the null energy condition).

In the case of quintessence scenario, and focusing on the basic model where
an exponential potential $V=V_0 e^{-\lambda\kappa\phi}$ for the scalar field
is imposed, introducing the auxiliary variables \cite{Copeland:1997et} 
\begin{equation}
x\equiv \frac{\kappa \dot\phi }{\sqrt{6}H}, \quad y\equiv \frac{\kappa \sqrt{%
V}}{\sqrt{3}H},  \label{auxbasicquint}
\end{equation}
we result to the dynamical system 
\begin{align}
& x^{\prime }=\frac{3}{2}x\left[ 2x^{2}+\gamma_{m}(1-x^{2}-y^{2})\right]
-3x+\sqrt{\frac{3}{2}}\lambda y^{2},  \label{eqxquint} \\
& y^{\prime }=\frac{3}{2}y\left[ 2x^{2}+\gamma_{m}(1-x^{2}-y^{2})\right] -%
\sqrt{\frac{3}{2}}\lambda xy,  \label{eqyquint}
\end{align}
with
$\gamma_m\equiv w_m+1$, 
in terms of which the various density parameters are expressed as $
\Omega_{d}= x^2+y^2$, $\Omega_m=1-\Omega_{d}$, while $w_{d}=\frac{x^2-y^2}{%
x^2+y^2}$. The critical points of the system \eqref{eqxquint}-%
\eqref{eqyquint}, along with their stability conditions and the
corresponding values of $\Omega_{d}$ and $w_{d}$ are shown in Table \ref%
{Quintback}. As we observe, the scenario possesses two stable late-time
attractors, with the scalar field dominated solution $D$ being the most
physically interesting.

\section{Dynamical analysis at the perturbation level} 
The investigation of
scalar perturbations is crucial in every cosmological scenario, since they
are connected to perturbation-related observables such as the growth index $%
\gamma$ and the $\sigma_8$ \cite{Bahcall:1998ur}. From now on, and for
calculation convenience, we focus on the most interesting case of dust
matter, namely we set $\gamma_{m}=1$ ($w_m=0$), since a 
non-zero $w_m$ does not qualitatively affect our results.

In a general non-interacting scenario, which includes dust matter and
dynamical dark energy, the scalar perturbations in the Newtonian gauge are
determined by the equations \cite{Ma:1995ey} 
\begin{eqnarray}
&&\dot{\delta}_m+\frac{\theta_m}{a}=0,\;  \label{eq:line1} \\
&&\dot{\delta}_{d}+(1+{{w_{d}})\frac{\theta_{d}}{a}+3H(c_{eff}^{2}-w_{d})%
\delta_{d}=0,\;}  \label{eq:line2} \\
&&\dot{\theta}_m+H\theta_m-\frac{k^{2}\psi }{a}=0,\;  \label{eq:line3} \\
&&\dot{\theta}_{d}+H\theta_{d}-\frac{k^{2}c_{\mathrm{eff}}^{2}\delta_{d}}{(1+%
{{w_{d}})a}}-\frac{k^{2}\psi }{a}=0,  \label{eq:line4}
\end{eqnarray}
where $k$ is the wavenumber of Fourier modes, and $\psi$ the scalar metric 
perturbation assuming zero anisotropic stress.
Additionally, $\delta_i\equiv \delta\rho_i/\rho_i$ are the densities
perturbations and $\theta_i$ are the velocity perturbations \cite{Ma:1995ey}. 
Furthermore,  $c_{\mathrm{eff}}^{2}$ is the effective sound
speed of the dark energy perturbations (the corresponding quantity for
matter is zero in the dust case), which determines the amount of dark-energy
clustering. Note that the above equations can be simplified by considering the Poisson 
equation, which in sub-horizon scales becomes \cite{Ma:1995ey}:
\begin{equation}
-\frac{k^{2}}{a^{2}}\psi =\frac{3}{2}H^{2}[\Omega_m\delta_{m}+(1+3c_{\mathrm{%
eff}}^{2})\Omega_{d}\delta_{d}]\;.  \label{eq:poisson}
\end{equation}
Finally, we 
mention that the above perturbation equations must be considered
alongside the background evolution equations (\ref{FR1})-(\ref{eqn:contdt}).

In  general, the fact that $\Lambda$ 
does  not  change  in  space
and time implies that the cosmological constant 
can  not  cluster  like  dark  matter.  
On the other hand, dynamical dark energy may cluster and the
amount of clustering is affected by the effective sound speed.
Specifically, in the case of $c_{\mathrm{eff}}^{2}=1$, pressure 
suppresses any dark energy fluctuation at sub-horizon scales. 
Therefore, for  
homogeneous dark energy the quantities $\delta_{d}$ and $\theta_{d}$ are  
vanished. On the other hand, for $c_{\mathrm{eff}}^{2}=0$
dark energy clusters similar to that of dark matter and perturbations
will grow with time. The clustering of dark energy modifies 
the evolution of dark matter fluctuations perturbations, hence 
it affects the structure formation rate of the universe (for more 
discussion see \cite{Mehrabi:2015hva} and references therein).

Let us first investigate the case of $\Lambda$CDM paradigm, which is
obtained by the above general framework for  $w_d=-1$ and $\rho_d=\Lambda/\kappa^2$, 
alongside $\delta_d=0$ and $\theta_d=0$ (i.e. dark energy is not clustering and thus 
its perturbation equations can be completely ignored). As auxiliary variables we
introduce $\Omega_m$, as well as the variable 
\begin{equation}
 {U}_m\equiv\frac{\delta_m^{\prime }}{\delta_m}.
\end{equation}
Hence, in terms of $\Omega_m$, $ {U}_m$, the equations (\ref{FR1})-(%
\ref{eqn:contdt}) and (\ref{eq:line1})-(\ref{eq:line4}) become 
\begin{eqnarray}  \label{LCDM-perts}
&&\Omega_m^{\prime }= 3 (\Omega_{m}-1)\Omega_{m}, \\
&&U_m^{\prime }= \frac{3}{2}(U_{m}+1) \Omega_{m}-U_m (U_m+2).
\label{LCDM-perts2}
\end{eqnarray}
The critical points of the system \eqref{LCDM-perts}-\eqref{LCDM-perts2},
along with the corresponding eigenvalues and their stability conditions are
presented in Table \ref{Lambdapert}. 
\begin{table}[ht]
\begin{center}
\begin{tabular}{|c|c|c|c|c|c|}
\hline
C.P. & $\Omega_m$ & $ {U}_m$ & Existence & Eigenvalues & Stability \\ 
\hline\hline
$P_1$ & 0 & $-2$ & Always & $\{-3,2\}$ & Saddle \\ \hline
$P_2$ & 1 & $-\frac{3}{2}$ & Always & $\left\{3,\frac{5}{2}\right\}$ & 
Unstable \\ \hline
$P_3$ & 0 & 0 & Always & $\{-3,-2\}$ & Stable \\ \hline
$P_4$ & 1 & 1 & Always & $\left\{3,-\frac{5}{2}\right\}$ & Saddle \\ \hline
\end{tabular}%
\end{center}
\caption[Lambdapert]{The critical points and their stability conditions, of both 
background and perturbation equations, in the case of $\Lambda$CDM paradigm.}
\label{Lambdapert}
\end{table}
The system admits four critical points, with $P_3$ being the stable one. It
corresponds to the cosmological-constant dominated, de Sitter solution,
which moreover has $\delta_m=\text{const.}$ (since $U_m=0$). Similarly, one
can observe the saddle point $P_4$, which is a matter dominated universe in
which the perturbations increase as $\delta_m\propto e^{ \tau}=a$ exactly at
the critical point. Thus, for $\Lambda$CDM cosmology the incorporation of
perturbations does not change the late-time attractor of the background
evolution.

For completeness we must examine the possibility of critical points that
exist at ``infinity'' and hence that are missed through the above basic
analysis. Introducing the transformation $\{\Omega_m,U_m\}\rightarrow\{%
\Omega_m, \bar{U}_m\}$ with $ \bar{U}_m=\frac{2}{\pi}\arctan(U_m)$, we find that such 
critical points at infinity do not exist, since    $\bar{U}_m'|_{\bar{U}_m=\pm 
1}=-2/\pi\neq 0$.

\begin{figure}[h]
    \centering
 \includegraphics[height=1.5in]{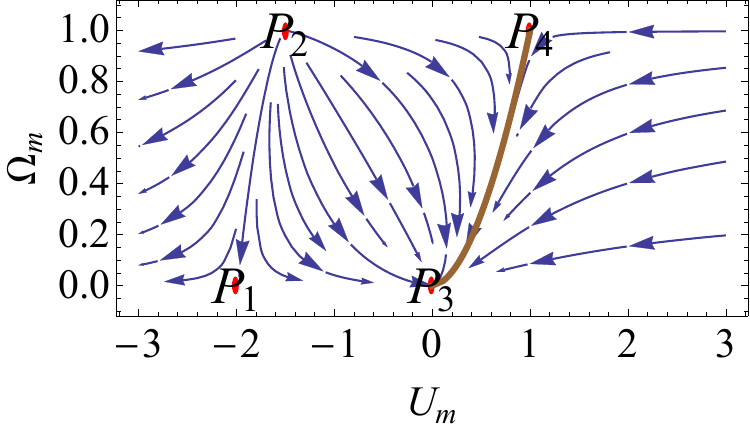}\\
    \caption{{\it{\label{fig:LCDM_perts} 
  The phase-space diagram for $\Lambda$CDM cosmology, at both background and 
perturbation levels. At late times the system is attracted by the de-Sitter point $P_3$. 
The thick line is  the curve (\ref{mattercurve}), which coincides with the unstable 
manifold of the matter dominated solution $P_4$, and which for  $\Omega_m$ close to 1
gives analytically  $\gamma=\frac{6}{11}$ as expected (see text).}}    
  }
\end{figure}

Finally, we note that in the literature it is standard to consider that in
the matter-dominated phase, in which the large scale structure builds up due
to the increase of matter perturbations, we have the relation $%
d\ln\!\delta_m/d\ln\! a\simeq\Omega_m^\gamma$ where $\gamma$ is the growth
index \cite{Peebles:1994xt}, which in our notation becomes just $%
U_m\simeq\Omega_m^\gamma$. Inserting it into \eqref{LCDM-perts2} we obtain 
\begin{equation}
3 \gamma (\Omega_m-1) \Omega_m^\gamma+\left(\Omega_m^\gamma+2\right)
\Omega_m^\gamma-\frac{3}{2}\Omega_m \left(\Omega_m^\gamma+1\right)=0,
\label{mattercurve}
\end{equation}
which expanded around $\Omega_m=1$ leads to 
\begin{equation}
-\left(\frac{11 \gamma}{2}-3\right) (1-\Omega_m)+{ {O}}%
\left((1-\Omega_m)^2\right)=0.
\end{equation}
As expected the asymptotic value of
the growth index is $\gamma=\frac{6}{11}$. 
%for the matter dominated universe as
%expected. 
The curve (\ref{mattercurve}) is depicted in Fig. \ref{fig:LCDM_perts} 
with a thick (brown) line, and as we observe it coincides with the unstable manifold of
the matter dominated solution $P_4$.

\begin{table*}[ht]
\begin{center}
\begin{tabular}{|c|c|c|c|c|c|c|c|}
\hline
$ {C.P.}$ & $ {\{x,y\}}$ & $ {U_{m} }$ & $%
 {\Omega }_{m}$ & $w_{d}$ & Existence & Eigenvalues &  Stability \\ \hline
\hline
$A_{1}$ & $\{0,0\}$ & $-\frac{3}{2}$ & $1$ & Undefined & Always & 
$\frac{5}{2},-\frac{3}{2},\frac{3}
{2}$ & Saddle
\\ \hline
$A_{2}$ & $\{0,0\}$ & $1$ & $1$ & Undefined & Always & 
$-\frac{5}{2},-\frac{3}{2},\frac{3}{2}$ & 
Saddle \\ \hline
$B_{1}$ & $\{1,0\}$ & $0$ & $0$ & $1$ & Always & $3,1,3-\sqrt{\frac{3}{2}} \lambda $ & 
Unstable for 
$\lambda <\sqrt{6}$ \\
  &   &  &   &  &   &   & Saddle for $\lambda >\sqrt{6}$ \\ \hline
$B_{2}$ & $\{1,0\}$ & $1$ & $0$ & $1$ & Always & $3,-1,3-\sqrt{\frac{3}{2}} \lambda$ & 
Saddle \\ \hline
$C_{1}$ & $\{-1,0\}$ & $0$ & $0$ & $1$ & Always & $3,1,3+\sqrt{\frac{3}{2}} \lambda$ & 
Unstable for 
$\lambda >-\sqrt{6}$ \\
  &   &  &   &  &   &   & Saddle for $\lambda <-\sqrt{6}$ \\ \hline
$C_{2}$ & $\{-1,0\}$ & $1$ & $0$ & $1$ & Always & $3,-1,\sqrt{\frac{3}{2}} \lambda +3$ & 
Saddle \\ \hline
$D_{1}$ & $\{\frac{\lambda }{\sqrt{6}}, \sqrt{1-\frac{\lambda ^{2}%
}{6}}\}$ & $0$ & $0$ & $-1+\frac{\lambda^2}{3}$ & $\lambda^2\leq 6$ & $\lambda 
^2-3,\frac{1}{2} \left(\lambda^2-6\right),
\frac{1}{2} \left(\lambda ^2-4\right)$ &  Stable for 
$\lambda^2<3$ \\
	& & & & & & & Saddle for  $\lambda^2>3$\\
\hline
$D_{2}$ & $\{\frac{\lambda }{\sqrt{6}}, \sqrt{1-\frac{\lambda ^{2}%
}{6}}\}$ & $\frac{\lambda ^{2}}{2}-2$ & $0$ & $-1+\frac{\lambda^2}{3}$ & $\lambda^2\leq 
6$ 
& $\lambda ^2-3,\frac{1}{2} \left(\lambda
   ^2-6\right),-\frac{1}{2} \left(\lambda ^2-4\right)$ & 
Saddle \\
\hline
$E_{1}$ & $\{\frac{\sqrt{\frac{3}{2}}}{\lambda }, \frac{\sqrt{%
\frac{3}{2}}}{\lambda }\}$ & $-\frac{1}{4}\left(1-\sqrt{25-\frac{72}{\lambda^2}}\right)$ & 
$1-\frac{
3}{\lambda ^{2}}$ &0 & $\lambda^2\geq 3$ & 
$-\frac{1}{2}\sqrt{25-\frac{72}{\lambda^2}}
,-\frac{3}{4} \left(1\pm \sqrt{\frac{24}{\lambda^2}-7}\right)$ &
Stable for  $\lambda^2>3$
\\
\hline
$E_{2}$ & $\{\frac{\sqrt{\frac{3}{2}}}{\lambda }, \frac{\sqrt{%
\frac{3}{2}}}{\lambda }\}$ & $-\frac{1}{4}\left(1+\sqrt{25-\frac{72}{\lambda^2}}\right)$ & 
$1-\frac{
3}{\lambda ^{2}}$
& $0$ & $\lambda^2\geq 3$ & $\frac{1}{2}\sqrt{25-\frac{72}{\lambda^2}},-\frac{3}{4} 
\left(1\pm \sqrt{\frac{24}{\lambda^2}-7}\right)$& Saddle \\\hline
\end{tabular}
\end{center}
\caption {\label{tab1b} 
The  physical (real with $0\leq\Omega_{m}\leq1$ and expanding)  critical points, their 
stability conditions, and their properties, of both 
background and perturbation equations, in the case of quintessence with 
exponential potential.  The stability conditions arise from the examination of the 
sign of the eigenvalues of the involved 
perturbation matrix.}
\end{table*}

We mention here that the dynamical system analysis is not needed for $\Lambda
$CDM  cosmology, since even including the perturbations the system remains
integrable. In particular, the general solution reads 
\begin{eqnarray}
&& \!\!\!\!\!\!\!\!\!\!\!\!\!\!\!\!\!\!\Omega_m(\tau)=\frac{\Omega_{m0}}{e^{3 \tau } (1- 
\Omega_{m0})+\Omega_{m0}}%
,  \label{exact0}
\\
&& \!\!\!\!\!\!\!\!\!\!\!\!\!\!\!\!\!\! U_m(\tau)= \Big\{2 \Omega_{m} (2
U_{m0}+3 \Omega_{m0}) \big( \Omega_{m0}^{2/3} g_0- \Omega_m^{2/3} g \big) 
\notag \\
&& \ \ \ \ \ +8 (1-\Omega_{m0})^{5/6} \Omega_{m0}^{2/3} \Big\}^{-1}  \notag
\\
&&\ \ \, \cdot \Big\{ 3 \Omega_{m} (2 U_{m0}+3 \Omega_{m0}) \big( %
\Omega_m^{2/3} g -\Omega_{m0}^{2/3} g_0 \big)  \notag \\
&&\ \ \ \ \ + 4 \Omega_m^{2/3} \Big[(2 U_{m0}+3 \Omega_{m0})
\left(1-\Omega_m \right)^{5/6}  \notag \\
&&\ \ \ \ \ \ \ \ \ \ \ \ \ \ \ \ -3 \Omega_{m0}^{2/3}\Omega_m^{1/3}(1-
\Omega_{m0})^{5/6} \Big] \Big\},  \label{exact}
\end{eqnarray}
where $g(\tau)=\!\, _2F_ 1\left[\frac{1}{6},\frac{2}{3},\frac{5}{3},
\Omega_m(\tau)\right] $ and $g_0=\!\, _2F_ 1\left[\frac{1}{6},\frac{2}{3},%
\frac{5}{3}, \Omega_{m0}\right]$, with $\Omega_{m0}$ and $U_{m0}$ the values
of $\Omega_m$ and $U_m$ at $\tau=0$ (i.e. at $a=1$). From the analytical
solutions (\ref{exact0}),(\ref{exact}) we can easily see that for $%
\tau\rightarrow\infty$ we have $\Omega_m\rightarrow0$ and $U_m\rightarrow0$,
i.e the system results to the de Sitter point $P_3$.

We now proceed to the investigation of perturbations in quintessence with exponential 
potential. As we mentioned above, this simple dark energy scenario has 
$c_{\mathrm{eff}}^{2}=1$, which implies that dark energy is non-clustering, and hence one 
should   consider only the perturbation equations
(\ref{eq:line1}) and (\ref{eq:line3}), alongside the background  
(\ref{FR1})-(\ref{eqn:contdt}) ones.
In order to transform them into autonomous form we  use the variables $x,y$ of
(\ref{auxbasicquint}), as well as the additional variable 
\begin{equation}
  {U}_m=\frac{\delta_m^{\prime }}{\delta_m}.
  \label{Umvar}
\end{equation}
Therefore,   the autonomous dynamical system consists of Eqs. 
\eqref{eqxquint}, \eqref{eqyquint} and 
\begin{equation}
U_m^{\prime }=-U_m^2-\frac{U_m}{2}
 \left(1-3x^2+3y^2\right)+\frac{3}{2}  \left(1-x^2-y^2\right),\label{eqxquintUM} 
\end{equation}
i.e. it is now 3-dimensional in contrast to the 2-dimensional one of the background 
equations. Since the first two equations are 
decoupled from the third one, the system admits the  five critical points of the 
background analysis of Table  \ref{Quintback},  
each of which is now split into two points due to the additional variable $U_m$. 
The  physical  critical points  and their stability conditions are presented in Table 
\ref{tab1b}.    Finally, the analysis at infinity shows that stable critical points do 
not 
exist.

The crucial feature, which lies in the center of the analysis of this work, is 
that the stability  and properties of the points changes, due to the existence of extra 
dimensions (reflecting the incorporation of perturbation equations) in the phase space. 
In particular, we can see that the only two points that can be conditionally 
stable are $D_1$ and $E_1$. For $E_1$ we have $U_m>0$, which implies 
that $\delta_m$ increases 
exponentially in an expanding universe, and hence is not physically interesting. 
For $D_1$, although we obtain $U_m=0$ this point is not physically interesting since it 
has  $c_d^2\equiv \frac{d p_d}{d 
\rho_d}=\frac{p_d'}{\rho_d'}=1+\sqrt{\frac{2}{3}}\lambda 
x(w_d-1)/(w_d+1)=-1+\frac{\lambda^2}{3}<0$ 
in its stability region, resulting to Laplacian instabilities.  
 Therefore, the incorporation of perturbation ruins the  
dark-energy dominated, de-Sitter solution, which is the physically interesting late-time 
attractor of the background equations, since it induces to it Laplacian instabilities.

\section{Conclusions}  
We performed for the first time 
a dynamical system analysis of both the background and 
perturbation equations, of $\Lambda$CDM cosmology and quintessence
 scenario with 
exponential potential. 
In the former case, 
the incorporation of perturbations does not change the 
stability of the late-time 
attractor of the background equations, and the system still results in the dark-energy 
dominated, de Sitter solution, having passed from the 
correct dark-matter era with 
$\gamma\approx 6/11$ (actually in this scenario one extracts analytical solutions). 
However, in the case of quintessence, the incorporation of perturbation changes the 
stability and properties of the background evolution, and the only conditionally stable 
points present either   an exponentially increasing matter 
clustering not favored by observations, or Laplacian instabilities, and thus   not 
physically interesting. In 
summary, the above results are a severe disadvantage of quintessence with 
exponential potential, (which is the archetype scenario due to the well-posed theoretical 
justification of exponential potentials) comparing 
to $\Lambda$CDM paradigm.

\end{document}